\documentclass[sigconf, nonacm]{acmart}

\newcommand\vldbyear{2025}
\newcommand\vldbworkshop{Applied AI for Database Systems and Applications (AIDB 2025)}
\newcommand\vldbauthors{\authors}
\newcommand\vldbtitle{\shorttitle} 
\newcommand\vldbavailabilityurl{https://github.com/DataManagementLab/JOB-Complex}
\newcommand\vldbpagestyle{plain} 

\usepackage{clev eref}
\usepackage{enumitem}
\usepackage[]{hyperref}
\usepackage{xurl}
\usepackage{enumitem}
\usepackage{multirow}
\usepackage{xspace}
\usepackage{cleveref}
\usepackage{todonotes}
\usepackage{listings}
\usepackage[htt]{hyphenat}
\usepackage[nolist]{acronym}

\usepackage{pifont}
\newcommand{\cmark}{\ding{51}}%
\newcommand{\benchmark}{\textsc{JOB-Complex}\xspace}

\begin{document}

\begin{acronym}
\acro{LCM}{Learned Cost Model}
\end{acronym}

\title{\benchmark: A Challenging Benchmark for\\Traditional \& Learned Query Optimization}

\author{Johannes Wehrstein}
\orcid{0000-0002-7152-8959}
\affiliation{\institution{Technical University of Darmstadt}}

\author{Timo Eckmann}
\orcid{0009-0007-7497-2389}
\affiliation{\institution{Technical University of Darmstadt}}

\author{Roman Heinrich}
\orcid{0000-0001-7321-9562}
\affiliation{\institution{Technical University of Darmstadt}\institution{DFKI}}

\author{Carsten Binnig}
\orcid{0000-0002-2744-7836}
\affiliation{\institution{Technical University of Darmstadt}\institution{DFKI}}

\begin{abstract}
Query optimization is a fundamental task in database systems that is crucial to providing high performance.
To evaluate learned and traditional optimizer's performance, several benchmarks, such as the widely used JOB benchmark, are used.
However, in this paper, we argue that existing benchmarks are inherently limited, as they do not reflect many real-world properties of query optimization, thus overstating the performance of both traditional and learned optimizers.
In fact, simple but realistic properties, such as joins over string columns or complex filter predicates, can drastically reduce the performance of existing query optimizers.
Thus, we introduce \benchmark, a new benchmark designed to challenge traditional and learned query optimizers by reflecting real-world complexity.
Overall, \benchmark contains 30 SQL queries and comes together with a plan-selection benchmark containing nearly 6000 execution plans, making it a valuable resource to evaluate the performance of query optimizers and cost models in real-world scenarios.
In our evaluation, we show that traditional and learned cost models struggle to achieve high performance on \benchmark, providing a runtime of up to 11x slower compared to the optimal plans.
\end{abstract}

\maketitle
\pagestyle{\vldbpagestyle}
\begingroup\small\noindent\raggedright\textbf{VLDB Workshop Reference Format:}\\
\vldbauthors. \vldbtitle. VLDB \vldbyear\ Workshop: \vldbworkshop.\\
\endgroup
\begingroup
\renewcommand\thefootnote{}\footnote{\noindent
This work is licensed under the Creative Commons BY-NC-ND 4.0 International License. Visit \url{https://creativecommons.org/licenses/by-nc-nd/4.0/} to view a copy of this license. For any use beyond those covered by this license, obtain permission by emailing \href{mailto:info@vldb.org}{info@vldb.org}. Copyright is held by the owner/author(s). Publication rights licensed to the VLDB Endowment. \\
\raggedright Proceedings of the VLDB Endowment. 
ISSN 2150-8097. \\
}\addtocounter{footnote}{-1}\endgroup

\ifdefempty{\vldbavailabilityurl}{}{
\vspace{.3cm}
\begingroup\small\noindent\raggedright\textbf{VLDB Workshop Artifact Availability:}\\
The source code, data, and/or other artifacts have been made available at \url{\vldbavailabilityurl}.
\endgroup
}

\section{Introduction}
\begin{figure}\centering
  \includegraphics[width=.9\linewidth]{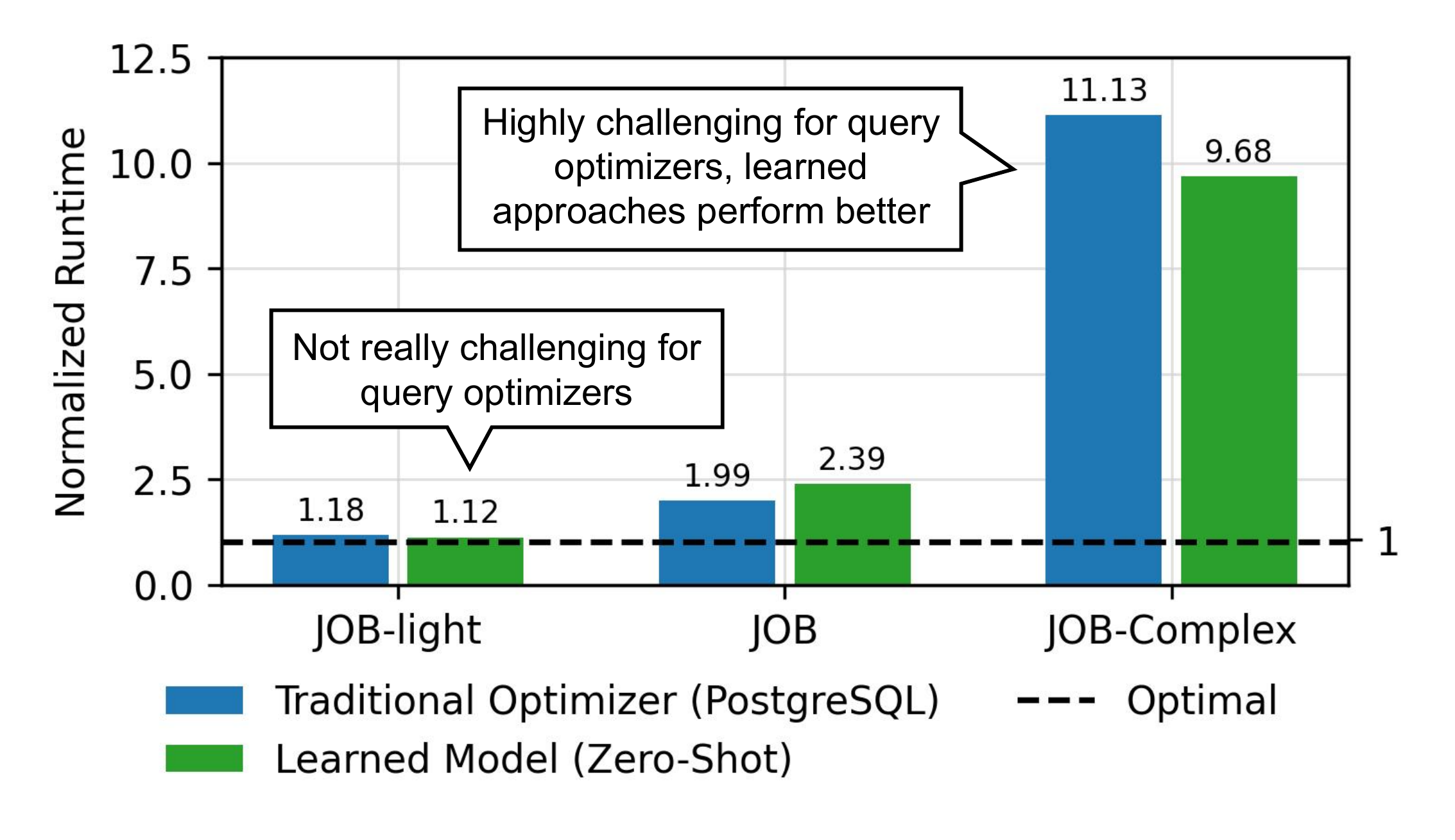} 
  \vspace{-3ex}
  \caption{Query optimization performance for PostgreSQL and a selected learned cost model Zero-Shot \cite{DBLP:journals/pvldb/HilprechtB22} over JOB-Light, JOB and \benchmark (ours). 
  Here, we report the sum of the selected plan runtimes, normalized by the optimal runtime (optimization gap).
  In contrast to existing benchmarks, \benchmark poses significant challenges for query optimization by showing an high optimization gap of up to 11.13.}
  \label{fig:optimization_gap_overview}
 \vspace{-5ex}
\end{figure}

\noindent\textbf{The State of Query Optimization}.
Efficient query optimization is the backbone of providing high-performance database management systems.
For a given SQL statement, the query optimizer has to choose optimally among many logical execution plans.
This is highly important, as a bad plan may take orders of magnitude longer than the fastest one \cite{leis2015, heinrich2025}.
Thus, query optimizers aim to minimize the query runtime, e.g., by enumerating different join orders or selecting between physical join operators.
However, query optimization is a highly challenging task, as the number of plan candidates grows quickly with the query complexity.
For that reason, query optimizers have been under constant improvement for decades \cite{jarke1984}.
While commercial database systems today typically apply static rule-based optimizations and hand-crafted cost models, \textit{machine learning} approaches have been proposed in recent years to further improve query optimization\cite{DBLP:conf/sigmod/WuMLNNPSRNK24,DBLP:conf/sigmod/NegiIMAKFJ21}, covering \acp{LCM} \cite{DBLP:journals/pvldb/HilprechtB22, DBLP:conf/icde/LiangCXYCX024, rieger2025}, learned cardinalities \cite{DBLP:journals/pvldb/HilprechtSKMKB20, wehrsteintowards, DBLP:journals/pvldb/YangKLLDCS20}, learned query hinting \cite{DBLP:journals/corr/abs-2004-03814}, and even end-to-end learned query optimizers\cite{DBLP:journals/pvldb/MarcusNMZAKPT19,DBLP:conf/sigmod/YangC0MLS22, DBLP:journals/pvldb/ZhuCDCPWZ23}.

\noindent\textbf{Is Query Optimization A Solved Problem (in times of AI)?}
To exercise the performance of query optimizers, many benchmarks were proposed, such as JOB-light \cite{DBLP:conf/cidr/KipfKRLBK19}, JOB \cite{leis2015} and TPC-H \cite{tpch}.
Particularly, JOB and JOB-light specifically evaluate the performance of query optimizers and cost models in choosing good join orders (hence the name Join-Order-Benchmark).
However, in this paper, we examined classical optimizers like PostgreSQL and learned optimizers such as ZeroShot \cite{DBLP:journals/pvldb/HilprechtB22} and our findings, and find, that these optimizers provide already almost optimal performance on these benchmarks, as shown in \Cref{fig:optimization_gap_overview}.
For instance, on JOB-light, PostgreSQL achieves to select plans only 1.18x slower than the optimal plan which we refer to as an \textit{optimization gap} of 1.18.
While JOB is significantly more complex than JOB-light, the overall performance is still remarkably high. 
Here, PostgreSQL chooses for 50\% of the queries a plan that is less than 29\% slower than the optimal plan.
On the whole benchmark, the optimization gap amounts to only 1.99, as shown in \Cref{fig:optimization_gap_overview}.
Similarly, \acp{LCM} like Zero-Shot \cite{DBLP:journals/pvldb/HilprechtB22} achieve a similar quality of plan selection with optimization gaps of 1.12 for JOB-light and 2.39 for JOB.
Hence, we want to raise the question in this paper:
\textit{Do we need to further invest in query optimization and learned models, or is query optimization an already solved problem in times of AI?}

\begin{table*}[]
\centering
\scalebox{0.9}{
\begin{tabular}{c|c|cccc|cc|c}
\hline
\begin{tabular}[c]{@{}c@{}}Benchmark\end{tabular} & \begin{tabular}[c]{@{}c@{}}Number of\\ Queries\end{tabular} & \begin{tabular}[c]{@{}c@{}}Number of\\Joins\end{tabular} & \begin{tabular}[c]{@{}c@{}}String\\ Filters\end{tabular} & \begin{tabular}[c]{@{}c@{}}Join on\\ Non-PK/FK\\ columns\end{tabular} & \begin{tabular}[c]{@{}c@{}}Join on\\ String Columns\end{tabular} & \multicolumn{1}{c}{\begin{tabular}[c]{@{}c@{}}Runtime of \\ Optimal Plans (s)\end{tabular}} & \multicolumn{1}{c|}{\begin{tabular}[c]{@{}c@{}}Runtime of \\ PG selected Plans (s)\end{tabular}} & \multicolumn{1}{c}{\begin{tabular}[c]{@{}c@{}}\textbf{Optimization} \\ \textbf{Potential}\end{tabular}} \\ \hline
JOB-light \cite{DBLP:conf/cidr/KipfKRLBK19} & 70 & 1-3 & -- &  -- &  -- & 2359.72 & 2795.53 & \textbf{1.18} \\
JOB \cite{leis2015} & 113 & 3-14 & \cmark &  -- &  -- & 156.79 & 312.23 & \textbf{1.99} \\
\benchmark (ours) & 30 & 5-14 & \cmark & \cmark & \cmark & 53.34 & 593.50 & \textbf{11.13} \\
\hline
\end{tabular}}
\caption{Key characteristics of existing benchmarks. 
\benchmark comes with real-world query properties that pose a significant challenge for query optimizers, as reflected in a very high runtime of selected plans by PostgreSQL, that is 11.13x slower than the optimal plans.
The runtimes are aggregated over all queries in each benchmark.}
\label{tab:benchmark_characteristics}
\vspace{-5ex}
\end{table*}

\textbf{The Need For A New Benchmark.}
While these standard benchmarks have been a valuable tool for evaluating query optimizers, our experience with them revealed that they lack many complex real-world challenges. 
As a result, these benchmarks (while valuable) do not effectively stress query optimizers or expose their limitations.
For instance, TPC-H uses synthetic data lacking real-world correlations and complex data distributions, often leading to overly accurate cardinality estimates, as recently discussed \cite{DBLP:journals/pvldb/RenenHPVDNLSKK24}.
JOB \& JOB-light were proposed to address this using the IMDB movie dataset, a real-world dataset with complex correlations.
While these benchmarks capture varying levels of query complexity, they, however, still largely operate under simplifying assumptions, such that data exists in a normalized schema or that primary/foreign key (PK/FK) constraints are rigidly enforced.
However, the scenarios faced by modern database systems often extend far beyond these assumptions \cite{DBLP:journals/pvldb/RenenHPVDNLSKK24,ding2021dsb,dreseler2020quantifying,leis2018query,negi2023unshackling}.
For instance, users frequently join on non-key columns \cite{ding2021dsb}, use string-based join conditions, or employ complex filter predicates involving \texttt{LIKE} or \texttt{IN} clauses with many values \cite{ding2021dsb,leis2018query, dreseler2020quantifying}. 
Moreover, data might not always be perfectly normalized, and schemas can evolve over time \cite{DBLP:journals/pvldb/RenenHPVDNLSKK24}.
However, current standard benchmarks fail to adequately capture these properties, limiting their effectiveness in evaluating query optimizers and cost models and often resulting in overly optimistic performance assessments.

\noindent\textbf{Introducing \benchmark.}
To address this gap of realistic benchmarks, we introduce \benchmark, a novel benchmark to evaluate query optimization and cost models.
\benchmark consists of 30 hand-crafted SQL queries that reflect real-world conditions, such as joins on string columns, joins on non-primary keys, and complex filter predicates.
Moreover, different from other benchmarks, \benchmark includes multiple execution plans for each query, encompassing physical plans along with their estimated and actual costs and cardinalities. 
This enables a reproducible and comparable evaluation of cost models and query optimizers, making \benchmark a valuable resource for benchmarking.
To create \benchmark, we adapted query patterns from the JOB benchmark, demonstrating how relatively simple modifications can result in a significantly more challenging and realistic query optimization benchmark. 
As shown in \Cref{fig:optimization_gap_overview}, \benchmark presents a much greater challenge for both traditional and learned approaches compared to JOB and JOB-Light, with optimization gaps of up to 11.13 for PostgreSQL and 9.68 for the learned ZeroShot model.

\noindent\textbf{Contributions.}
Overall, we present the following contributions in this paper:
\begin{enumerate}[leftmargin=*, nosep]
    \item We introduce \benchmark\footnote{https://github.com/DataManagementLab/JOB-Extended}, a novel benchmark for query optimization and cost estimation that incorporates real-world challenges such as joins on non-primary/foreign key columns, joins on string columns, and complex filter predicates.
    \item We additionally provide a plan selection benchmark that covers a broad range of pre-executed plans for each SQL query of \benchmark, as well as JOB and JOB-light. 
    This enables better training of learned approaches and fine-grained evaluation of plan selection.
    \item In our experimental section, we show that both traditional and learned query optimizers and cost models struggle with the increased complexity of \benchmark, indicating considerable room for improvement. Through a detailed evaluation of PostgreSQL and several recent learned cost models, we highlight their respective strengths and weaknesses when applied to these more challenging queries.
    \item We make \benchmark publicly available to foster future research and development in query optimization, encouraging the community to address these identified challenges.
\end{enumerate}

\noindent\textbf{Outline}.
In the remainder of this paper, we first present the design ideas and specific modifications made to create \benchmark in \Cref{sec:benchmark}.
We then present a comprehensive evaluation of traditional and learned query optimization techniques on this new benchmark in \Cref{sec:experiments}. Finally, we conclude our work and outline future research directions in \Cref{sec:conclusion}.


\section{\benchmark}\label{sec:benchmark}
In the following, we first present \benchmark, which is the core contribution of this work in \Cref{subsec:design}, before we discuss the necessity of plan enumeration datasets in \Cref{subsec:need_for_query_plans}.

\subsection{Benchmark Design}\label{subsec:design}
In this section, we present the core ideas of \benchmark and discuss how it differs from previous benchmarks by representing real-world query optimization challenges, as summarized in \Cref{tab:benchmark_characteristics}.
Overall, our benchmark comes with 30 queries that have between 5-14 joins.
Moreover, it contains various real-world properties, that we discuss in the following.
By that, it shows an optimization gap of 11.13, posing way harder challenges for query optimization than existing benchmarks.

\noindent\textbf{Real World Properties.}
To design a novel benchmark that presents a significantly greater challenge for query optimizers than existing ones, we build on the established JOB benchmark \cite{leis2015}, addressing its simplifying assumptions through targeted query modifications. 
This approach enables a meaningful comparison between \benchmark and JOB by preserving similar query structures, thereby avoiding superficial changes such as only increasing the number of joins. 
Different from JOB, however, \benchmark incorporates real-world query conditions, as detailed in the following:

\begin{enumerate}[leftmargin=*, nosep]
    \item \textbf{Joins on Non-Primary/Foreign Key Columns:} 
    Unlike JOB and JOB-Light, which primarily use joins on columns with PK/FK relationships, \benchmark includes queries where joins are performed on columns that do not have such constraints. 
    This is a common pattern in real-world analytical queries or when integrating denormalized data. 
    In fact, the absence of PK/FK relationships makes cardinality estimation for these joins significantly harder, as the estimator cannot rely on uniqueness or referential integrity assumptions.
    
    \item \textbf{Joins on String Columns:}
    Existing benchmarks predominantly use integer columns for join conditions. 
    In contrast, \benchmark introduces joins on string columns. 
    String comparisons are computationally more expensive, and the distribution of string values can be more complex and harder to summarize than numerical data, posing additional challenges for selectivity estimation and cost estimation.

    \item \textbf{Comparisons Between Columns of Same Table:}
    In \benchmark, we intentionally added comparisons between col\-umns within the same table, e.g., \texttt{cn.name\_pcode\_nf = cn.name\_pcode\_sf}.
    Such appear in data validation checks or complex business logic e.g., start-time before end-time.
    This is particularly challenging for query optimizers because they typically lack statistics on intra-row correlations and assume column independence, leading to inaccurate selectivity and cardinality estimates. 
    Additionally, such comparisons limit the use of common optimization techniques like index usage or predicate push-down.
    
    \item \textbf{Complex Filter Predicates:} 
    Similar to JOB, we incorporated more complex filter attributes, such as \texttt{LIKE} predicates for pattern matching on strings and \texttt{IN} clauses with multiple values. 
    Such filters are often simplified or underrepresented in benchmarks.
    Estimating the selectivity of such predicates accurately is a highly challenging problem.
    
    \item \textbf{Preservation of Read-Sets and Correlations:} 
    We intentionally kept the set of tables accessed by a query (read-sets) from the original JOB queries. 
    The IMDB dataset underlying JOB is known for its real-world data distributions and high cross-table correlations. 
    That way, \benchmark inherits this inherent data complexity, ensuring that the new challenges are layered on top of an already non-trivial data landscape.
\end{enumerate}

\noindent \textbf{Example Query.}
In the following, we present an example query (Nr. 12) from \benchmark, that covers most of those key ideas from \benchmark that were previously discussed.

\lstset{escapeinside={<@}{@>}}
\begin{lstlisting}[language=SQL, basicstyle=\ttfamily\small, keywordstyle=\color{black}\bfseries, commentstyle=\itshape, showstringspaces=false, breaklines=true]
SELECT MIN(chn.name), ... 
FROM complete_cast cc, comp_cast_type cct1, comp_cast_type cct2, ... -- (11 other tables)
WHERE cct1.kind = 'cast' 
AND cct2.kind LIKE '%complete%' 
AND chn.name IS NOT NULL 
AND (chn.name LIKE '%man%' 
    OR chn.name LIKE '%Man%')   <@\textcolor{orange}{ -- Complex Predicates}@>
AND k.keyword IN (...) 
AND ... -- (other filters)
AND chn.id = ci.person_role_id          <@\textcolor{orange}{ -- w/o PK}@>
AND ak.name_pcode_cf=n.name_pcode_cf    <@\textcolor{orange}{ -- on strings}@>
AND ak.name_pcode_nf=chn.name_pcode_nf  <@\textcolor{orange}{ -- on strings}@>
AND ... -- (other join conditions)
\end{lstlisting}

\begin{figure}
  \includegraphics[width=\linewidth]{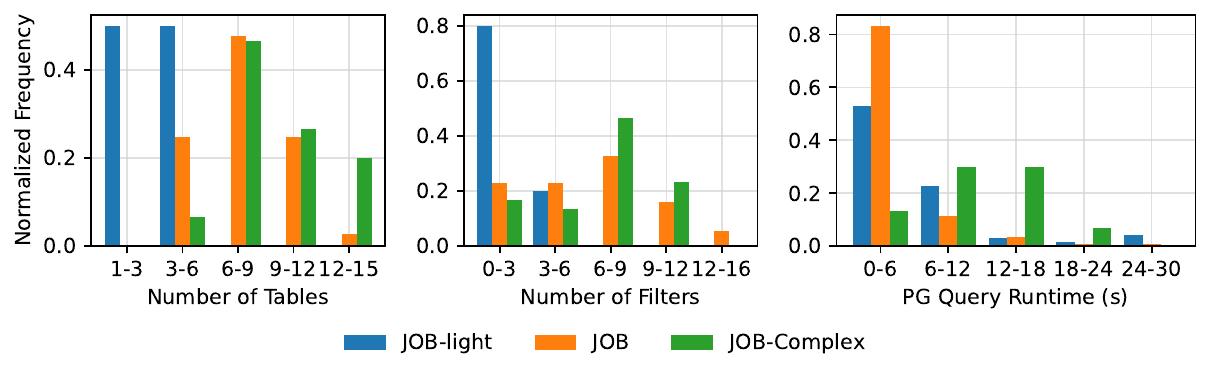} 
   \vspace{-3ex}
  \caption{Normalized distribution of number of tables and filter predicates per query across benchmarks.
  \benchmark is similar in the number of tables and filters to existing benchmarks.
  However, JOB-light \& JOB both show a skewed distribution of PostgreSQL runtimes of the queries, while \benchmark shows a more symmetric runtime distribution covering both short and long-running queries}
  \label{fig:num_tables_filters_hist}
  \vspace{-3ex}
\end{figure}

\noindent\textbf{Benchmark Design.}
To construct \benchmark, we selected 30 representative queries from the JOB benchmark and systematically applied the modifications described above. 
For instance, an original JOB query that joins on \texttt{A.id = B.fk\_id} (where \texttt{id} is a primary key and \texttt{fk\_id} is a foreign key) is transformed in \benchmark to join on columns such as \texttt{A.name = B.title} (string columns without key constraints) or \texttt{A.value1 = B.value2} (non-key numerical columns). 
This preserves the overall SQL complexity by maintaining similar numbers of joins, filters, and aggregations while introducing more realistic and challenging join and filter conditions.
By focusing on these targeted modifications, \benchmark enables a direct evaluation of how query optimizers cope with real-world complexities rather than simply increasing query size or structural complexity. 
As illustrated in \Cref{fig:num_tables_filters_hist}, the distribution of tables and filter predicates per query in \benchmark closely mirrors that of JOB, which importantly enables better comparisons. 
This alignment ensures that any observed differences in optimizer performance are due to the added real-world complexities in \benchmark rather than differences in query structure or size. 

Additionally, unlike JOB-light, all queries in \benchmark are guaranteed to produce non-empty results, preventing optimizers from exploiting shortcuts based on empty intermediate results.

\subsection{The Need for Query Plans}\label{subsec:need_for_query_plans}
Evaluating \acp{LCM} and query optimization techniques \emph{end-to-end} requires far more than a single plan per query.  
In reality, in query optimization scenarios, cost-models will be applied on a large set of query plan candidates, using their cost estimates to select the presumably fastest plan. 
Hence, to evaluate if a cost model's cost estimation leads to the correct ranking and selection of plans, a \textbf{diverse set of alternative execution plans} along with their actual execution costs is needed. 
Therefore, along the 30 SQL queries of \benchmark we also provide a meaningful selection of (on average) around 200 enumerated plans per SQL-query as the second key contribution of this work.
Providing enumerated and pre-executed plans allows developers of learned cost models and query optimizers to directly apply their cost models, thereby enabling the evaluation of query optimization and plan selection performance on \benchmark without the need to manually explore the search space for possible plans or construct a diverse set of plan candidates.

\noindent\textbf{Challenges in Plan Enumeration.}
However, generating a diverse and representative set of plans is challenging for multiple reasons:
\begin{enumerate}[leftmargin=*, nosep]
    \item The exponential growth of the search space with the number of joins. This makes the enumeration of all plans computationally infeasible for queries with more than a few joins.
    \item The tendency of naive enumeration to produce a skewed runtime distribution, where most plans are either extremely slow or very similar in performance.
    \item The lack of existing strategies for generating plans with 15+ joins while maintaining a uniform runtime distribution as these queries amplify both previous issues.
\end{enumerate}

\noindent\textbf{Proposed Solution.}
To address these challenges, we developed a steered plan generation procedure that ensures a diverse and balanced set of plans:

\noindent \textit{(1) Oracle-Guided Plan Generation}: 
We pre-calculated all intermediate cardinalities for all possible join orders (where feasible within a 30-second timeout per sub-plan/join-pair). 
These actual cardinalities were then injected into the PostgreSQL optimizer by modifying its estimates using \cite{pg_hint_plan}. 
This process helps to identify very good plans, as it mitigates the uncertainty introduced by inaccurate cardinality estimations that are provided by the default optimizer. 
That way, this strategy aims to find plans close to the true optimal.

\noindent \textit{(2) Diversified Random Enumeration}: 
Since the oracle-guided approach primarily finds good plans, it alone is not sufficient to explore suboptimal but plausible alternatives.
Thus, we also randomly enumerated join orders and enforced them in the PostgreSQL optimizer.
Though simply enumerating all possible join orders is computationally prohibitive due to exponential growth with the number of joins. 
In addition, a purely random enumeration often leads to a vast majority of plans being extremely slow and only a few achieving moderate or fast runtimes.
To address this problem, we propose a strategy where we only enforce a \textit{global join order} using \cite{pg_hint_plan} (i.e., the top-level sequence of joins, e.g., A $\bowtie $ B $\bowtie $ C) but allow the PostgreSQL optimizer to choose the best join \textit{nesting} (e.g., ((A $\bowtie $ B) $\bowtie$ C) vs. (A $\bowtie $(B$\bowtie $C)) if the global order allows for it) and physical operators for that enforced global order.
Further, we considered only tables with more than ten thousand rows in the global join order.
Smaller tables are omitted since they overall do not influence plan runtime significantly but drastically increase the number of possible join orders. 
This approach ensures a diverse distribution of overall plan structures while leveraging PostgreSQL's capabilities to optimize local decisions.
This approach avoids an overabundance of extremely poor plans and yields a more informative, diverse set of alternatives with a broad runtime spread. 
By combining oracle-guided and randomized enumeration, we provide, for each query, a representative set of plans (on average 200 per SQL query), enabling realistic and reproducible evaluation of cost models and plan selection strategies.

\section{Initial Results Using \benchmark}
\label{sec:experiments}
In this section, we use \benchmark to stress-test different cost models by comparing the performance of the PostgreSQL cost model for plan selection and several state-of-the-art learned cost models on \benchmark against established benchmarks like JOB and JOB-Light. 
In particular, we analyze (1) the overall query optimization performance, (2) the performance of traditional, (3) and learned plan selection and (4) cardinality estimation accuracy.

\noindent\textbf{Experimental Setup.}
All queries are executed on PostgreSQL v16 on instances of the academic cloud provider \textit{CloudLab} \cite{duplyakinRMWDES19}.
For the cost models, in addition to the cost model of PostgreSQL(16) we included a large variety of learned approaches including: DACE \cite{DBLP:conf/icde/LiangCXYCX024}, ZeroShot \cite{DBLP:journals/pvldb/HilprechtB22}, QPPNet \cite{DBLP:journals/pvldb/MarcusP19}, T3 \cite{rieger2025}, FlatVector \cite{DBLP:conf/icde/GanapathiKDWFJP09}, E2E \cite{DBLP:journals/pvldb/SunL19}, QueryFormer \cite{DBLP:journals/pvldb/ZhaoCSM22}, the cost-model of NEO \cite{DBLP:journals/pvldb/MarcusNMZAKPT19} and MSCN \cite{DBLP:conf/cidr/KipfKRLBK19}.
For the training procedures, we follow those described in \cite{heinrich2025}, to ensure a fair comparison by training the models all on the same datasets as much as possible.
Database-specific models are trained on the IMDB dataset that is underlying JOB / JOB-light and \benchmark, while database-agnostic models were pre-trained on a diverse set of 19 different databases, similar to \cite{DBLP:journals/pvldb/HilprechtB22}.

\subsection{Query Optimization Performance}\label{subsec:qo_perf}
\begin{figure}\centering
  \includegraphics[width=\linewidth]{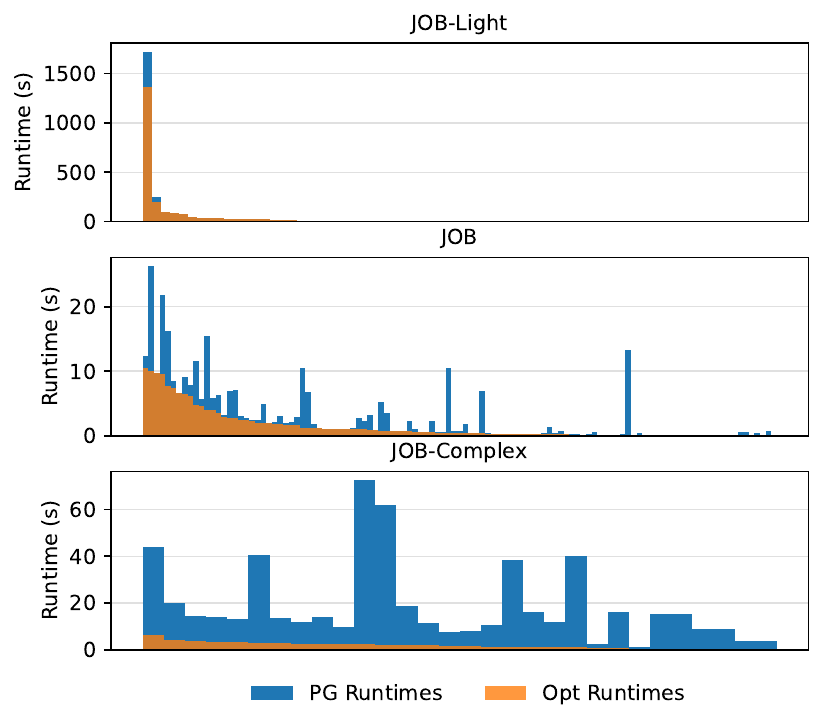} 
  \vspace{-3ex}
  \caption{Optimal runtime vs. PostgreSQL selected plan runtime per query of JOB-Light, JOB and \benchmark, sorted in descending runtime order. Overall, queries provided in \benchmark come with a high optimization potential.}
  \vspace{-2ex}
  \label{fig:speedups}
\end{figure}

Our initial analysis reveals that \benchmark poses a significantly greater challenge to traditional query optimization than established benchmarks like JOB and JOB-Light. 
This increased difficulty is visible in the substantially larger \textit{optimization gap} observed with \benchmark.
\Cref{fig:speedups} illustrates this gap by comparing the runtime of the plan selected by PostgreSQL against the runtime of the optimal plan (from our enumerated set) for each query, sorted by the optimal runtime. 
A higher optimization gap denotes a greater difficulty for the optimizer in finding the best plan.

The reasons for this larger gap on \benchmark are twofold:
First, both the absolute and relative differences between the optimal plan runtimes and the PostgreSQL-selected plan runtimes are considerably larger for \benchmark. 
This holds true across the entire spectrum of query execution times. 
For example, even for queries where the optimal plan executes in under 1 second, PostgreSQL frequently chooses plans on \benchmark that are 10 seconds or slower.
Second, \benchmark features a more uniform runtime distribution, unlike the long-tailed distributions of JOB and JOB-Light, where a few long running queries dominate. 
This reduces bias and ensures optimizers are tested consistently across both short and long-running queries, making \benchmark a more robust challenge.

These findings underscore that query optimization is far from a solved problem, particularly when faced with the real-world complexities captured by \benchmark. 
This naturally leads to the question: 
Why does PostgreSQL struggle to find optimal, or even near-optimal, plans on \benchmark? 
Is this primarily due to inaccuracies in its cardinality estimations or perhaps limitations within its cost model when dealing with these complex queries? 
And how do \acp{LCM} perform in more challenging scenarios?
To examine these questions, the following sections will analyze the accuracy of plan selection, predicted costs and cardinality estimates.

\subsection{How do Traditional Approaches Perform?}
\begin{figure}\centering
  \includegraphics[width=.95\linewidth]{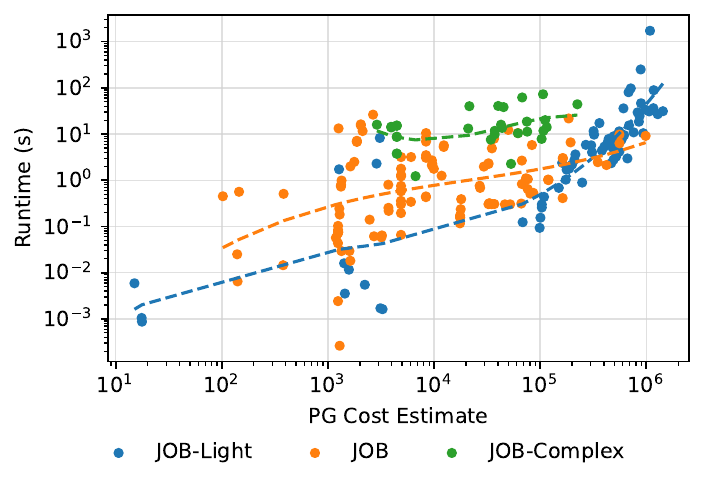} 
  \caption{PostgreSQL's estimated cost vs. actual runtime for queries in \benchmark. 
  Each point represents a query plan. 
  The dispersion indicates challenges in accurately modeling costs for the complex queries of \benchmark. 
  Ideally, points should lie close to a line with a positive slope, indicating a linear relationship.}
  \label{fig:cost_estimation} 
\end{figure}

In the following, we take a deeper look to understand why \benchmark stresses PostgreSQL more.
For this, we first analyze the accuracy of cost estimates as understanding the reliability of these estimates on complex queries is vital.
\Cref{fig:cost_estimation} shows PostgreSQL's estimated cost against the actual runtime for enumerated plans from \benchmark.
For JOB and JOB-Light, there is a tight correlation where estimated costs tend to increase with actual runtimes.
However, on \benchmark, this correlation is much weaker. 
While \benchmark has a more narrow runtime distribution (a desirable property, as discussed), the cost estimates from PostgreSQL vary by orders of magnitude and show little correlation with actual runtimes. 
This underscores the difficulty traditional cost models face in capturing the complexities introduced by \benchmark. 
Their struggle is also reflected in the median Q-error of the cost estimation by PostgreSQL as it is substantially higher on \benchmark in comparison to JOB-light and JOB as can be seen in  \Cref{tab:cost_est_qerror} (1015.67 on JOB, 3.87 on JOB-light and 2669.22 on \benchmark).

However, the question about the overall query optimization performance is not answered simply by cost estimation accuracy, as shown by \cite{heinrich2025}. 
Therefore, we analyze the optimization potential (selected plan runtime divided by optimal plan runtime) for PostgreSQL as well. 
On JOB, PostgreSQL achieves an impressively small optimization gap of roughly 2x, as previously shown in \Cref{fig:optimization_gap_overview} and \Cref{fig:speedups}.
In stark contrast, the same experiment on \benchmark reveals that traditional approaches do not consistently find near-optimal plans as the optimization gap is 11x. 
Since AI models have shown promising results for query optimization in recent years, we will in the following analyze if \acp{LCM} show similar struggles.

\subsection{How do AI Approaches Perform?}
To evaluate whether AI models show similar or different challenges, we selected nine representative LCMs, encompassing various architectures (GNNs, Transformers, set-based models, etc.) and approaches (database-specific and database-agnostic/zero-shot models like T3, ZeroShot and DACE).
Following the same steps to analyze the performance of traditional approaches, we first analyze the median Q-error of cost estimations before discussing the overall optimization gap of \acp{LCM}.

When examining the median Q-error of cost estimations for all models computed across all enumerated plans, including optimal and sub-optimal plans, as shown in \Cref{tab:cost_est_qerror}, an interesting pattern emerges:
Despite \benchmark proving significantly harder for plan selection (i.e., a larger optimization gap as shown in \Cref{subsec:qo_perf}), the median Q-errors for cost estimation by LCMs are often comparable or even slightly better on \benchmark than on JOB.
While accurate cost estimates are desirable, the ability to rank plans correctly for selection is essential. 

\begin{figure}
  \includegraphics[width=\linewidth]{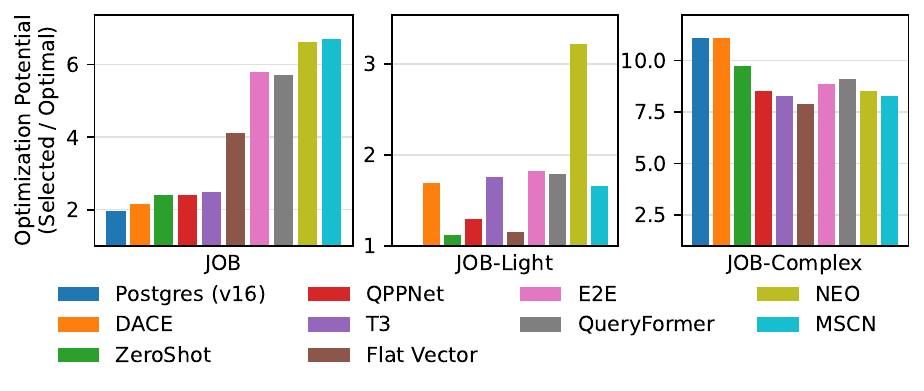}
  \vspace{-2ex}
  \caption{Optimization Potential (selected plan runtime / optimal plan runtime) for PostgreSQL (PG) and various \acp{LCM} across JOB-Light, JOB, and \benchmark. Higher values indicate poorer cost model performance and greater room for improvement. 
  Overall, \benchmark clearly shows a higher optimization potential across all models.}
  \label{fig:lcm_opt_potential}
  \vspace{-3ex}
\end{figure}

\begin{table}[]
\begin{tabular}{c|r|r|r}
\textbf{Median Q-Error} & \textbf{JOB} & \textbf{JOB-light} & \textbf{\benchmark} \\\hline\hline
Postgres (v16) & 1015.67 & 3.87 & 2669.22 \\\hline
DACE & 1.91 & 2.34 & 1.81 \\\hline
E2E & 13.21 & 5.03 & 23.39 \\\hline
FlatVector & 2.15 & 1.72 & 1.52 \\\hline
MSCN & 13.85 & 3.38 & 26.56\\\hline
NEO & 13.20 & 2.22 & 4.09 \\\hline
QPPNet & 17.50 & 1.45 & 67.20 \\\hline
QueryFormer & 8.30 & 3.49 & 9.88 \\\hline
T3 & 4.07 & 1.72 & 2.30 \\\hline
ZeroShot & 1.58 & 1.42 & 1.60 \\\hline\hline
\textbf{Median of all Models} & \textbf{10.75} & \textbf{2.28} & \textbf{6.99}
\end{tabular}
\caption{Cost Estimation Median Q-Error for Postgres and LCMs on all enumerated plan selection candidates. The increased difficulty in \benchmark significantly impacts cost estimation accuracy of PostgreSQL. However, the learned cost models show to be more robust against more complex queries and exhibit a roughly similar Q-error range than JOB.}
\label{tab:cost_est_qerror}
\end{table}
\textbf{What about plan selection?}
Next, we examine how challenging \benchmark is for \acp{LCM} when tasked with selecting the best execution plan from the enumerated set based on their cost estimates for each plan.
The results in \Cref{fig:lcm_opt_potential} show the previously discussed optimization gap for PostgreSQL against the selected \acp{LCM} across the three benchmarks.
On JOB, most advanced \acp{LCM} perform on par with PostgreSQL, showing an optimization potential of around 2 to 2.5. 
However, consistent with findings in \cite{heinrich2025}, most models, such as E2E, QueryFormer, NEO, and MSCN, show performance challenges.
On JOB-Light, PostgreSQL and several LCMs (e.g., ZeroShot, QPPNet, FlatVector) achieve near-optimal performance, with a significantly lower overall optimization potential, while other models fail to provide good plan selections.
In stark contrast to this optimization potential of roughly 2 to 2.5 on JOB, the optimization potential of the best-performing LCMs, such as T3 or FlatVector is roughly 8x on \benchmark. 
Clearly, this gap of 8x compared to the optimal is far from acceptable and highlights that query optimization is not solved, and there is a pressing need for improved cost models to tackle real-world query complexities. 
However, LCMs seem to perform better than traditional approaches on the more challenging \benchmark benchmark (gap of 8x for T3 and FlatVector vs. 11x for Postgres).
Motivated by \cite{leis2015} showcasing the huge impact of cardinality estimates on query optimization performance, we investigate in the following whether the increased complexity of \benchmark can also be attributed to cardinality estimation challenges.


\subsection{How Accurate are Cardinalities?}
\begin{figure}\centering
  \includegraphics[width=\linewidth]{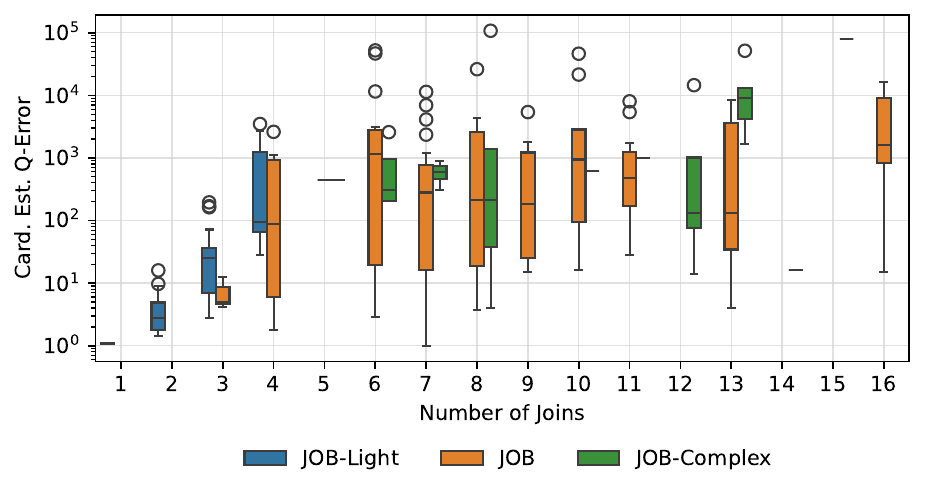} 
  \caption{Cardinality estimation error of PostgreSQL (q-error) at the root operator versus the number of tables in the query, compared across JOB-Light, JOB, and \benchmark. 
  \benchmark consistently exhibits higher q-errors, indicating less accurate cardinality estimates, especially as query complexity increases.}
  \label{fig:cardinality_estimation}
\end{figure}

Finally, we investigate the accuracy of cardinality estimates, which are fundamental inputs for cost models\cite{leis2015}, including PostgreSQL's native model and many \acp{LCM} (especially zero-shot models).
This experiment aims to determine whether the large optimization gap observed on \benchmark is primarily due to inaccurate cardinality estimates, or if the cost models themselves are unable to translate even reasonable cardinality estimates into good plan choices. 
By analyzing the cardinality estimation errors, we can assess whether improving cardinality alone would be sufficient, or if further improvements in cost modeling are necessary.
\Cref{fig:cardinality_estimation} compares the Q-error of PostgreSQL's cardinality estimates at the root operator, broken down by the number of joins in the query, across the three benchmarks.
The results show that while cardinality estimates for \benchmark are generally worse (higher Q-error) than for JOB-Light, they are in a somewhat similar range to those for JOB. 
As expected, JOB-Light, with its simpler queries and fewer complex correlations, yields better cardinality estimates.
The fact that the cardinality estimation errors on \benchmark are not drastically worse than on JOB, despite the introduction of challenging features like joins on non-key and string columns (notorious for causing estimation difficulties), is significant. 
It suggests that the large optimization gap observed on \benchmark does not solely stem from catastrophically worse cardinality estimates, as often suggested \cite{leis2015}.
Instead, it points towards inherent difficulties within the cost models themselves -- both traditional and learned -- in generalizing to and accurately costing more complex query patterns and operator inter-dependencies, even when underlying cardinality estimates are not excessively erroneous.
This indicates that improvements are needed not just rooted in cardinality estimation but also in how these estimates (and other plan features) are used by cost functions to predict final plan costs and guide plan selection.

\section{Conclusion and Future Work}\label{sec:conclusion}
In this work, we demonstrated that query optimization, particularly for complex Select-Project-Aggregate-Join (SPAJ) queries covering real-world properties, is far from a solved problem.
While existing benchmarks like JOB and JOB-Light suggest that systems like PostgreSQL or query optimization built on learned cost models achieve acceptable or even near-optimal query optimization performance, our findings indicate that this is an incomplete picture.
Thus, we introduced \benchmark, a novel benchmark derived from JOB, specifically designed to incorporate these real-world challenges.
By introducing realistic complexities, such as joins on non-key columns, string-based joins, and intricate filter predicates, we have shown that the optimization gap can quickly exceed 10x.
On \benchmark, PostgreSQL exhibits an 11x average optimization gap, and even state-of-the-art \acp{LCM} show a significant gap of around 8x.
Thus, our results highlight substantial room for improvement in both traditional, cost-model-based query optimizers and emerging learned approaches.
Interestingly, while LCMs struggle with the increased complexity of \benchmark as well, they tend to outperform traditional cost models.
This suggests that LCMs might possess a competitive advantage in handling more intricate query patterns, which, however requires further investigation.
Nevertheless, both paradigms are far from achieving optimal performance on these more challenging queries, signaling a clear need for new research into more robust and accurate cost estimation and plan selection techniques.


\noindent\textbf{Future Work.}
Our work represents a first step towards developing more realistic and challenging query optimization benchmarks.
Thus, it opens up several promising directions for future research.
\begin{enumerate}[leftmargin=*, nosep]
    \item \textbf{Increase Complexity of \benchmark:}
    The first direction is to extend \benchmark to cover an even broader range of real-world complexities, such as queries involving materialized views, semi-structured or nested data (e.g., JSON, XML), user-defined functions, and a wider spectrum of analytical query patterns (e.g., window functions, common table expressions).
    \item \textbf{Analyze Failures of Cost Models:} 
    Second, we aim to conduct a deeper root cause analysis for the suboptimal plans selected by traditional and learned approaches on \benchmark to better understand \textit{why} they selected those plans.
    However, this is particularly challenging for learned approaches, as they leverage deep neural networks and thus are not yet \textit{explainable}.
    \item \textbf{Advance Learned Approaches:} 
    Next, we see a large potential in learned approaches, as they have proven to provide highly accurate estimates in general.
    Thus, we suggest to further investigate the development of more sophisticated learned models, covering learned query plan representation, learned plan searching, and hybrid approaches that strategically combine the strengths of traditional optimizers with learned components.
    By providing various plan enumerations for the same queries in \benchmark, we importantly enable the development of such approaches.
    \item \textbf{Evaluating Query Optimization Across Systems:} 
    Moreover, we propose to evaluate \benchmark and its future extensions across a wider range of database systems, including open-source and commercial DBMSs, to validate the generality of our findings and identify system-specific strengths, weaknesses, and optimization strategies.
    \item \textbf{Generate Hard Queries Automatically:} 
    Finally, the creation of challenging benchmarks is a non-trivial task that requires expert knowledge and testing. 
    Thus, we propose to explore automated techniques for identifying and generating "hard" queries that maximally stress-test query optimizers beyond current benchmarks, which could be addressed with the help of LLMs, adversarial generation, or evolutionary algorithms.
\end{enumerate}

\noindent 
Ultimately, we aim to contribute to the development of more robust, effective, and truly general query optimizers capable of handling the full spectrum and complexity of real-world workloads.

\begin{acks}
This work has been supported by the LOEWE program (Reference III 5-519/05.00.003-(0005)), hessian.AI at TU Darmstadt, as well as DFKI Darmstadt. 
\end{acks}

\balance
\bibliographystyle{ACM-Reference-Format}
\bibliography{main}

\end{document}